\documentclass[superscriptaddress,twocolumn,prl]{revtex4-1}
\usepackage[latin9]{inputenc}
\usepackage{amsmath}
\usepackage{upgreek}
\usepackage{amstext}
\usepackage{graphicx}  
\usepackage{changes}
\usepackage{array}
\usepackage{subcaption}

\begin{document}
\title{A Search for Ultra-Light Vector Dark Matter with a Rotating Torsion Balance}
\author{M.P. Ross}
\author{ E.A. Shaw}
\author{ C. Gettings}
\author{ S.K. Apple}
\author{I.A. Paulson}
\author{ J.H. Gundlach}
\affiliation{Center for Experimental Nuclear Physics and Astrophysics, University of Washington, Seattle, Washington 98195, USA}

\begin{abstract}
We search for ultra-light vector dark matter interacting with a rotating torsion balance with a baryon minus lepton number composition dipole. Our search spans candidate masses in the ultra-low mass range from 1.3~$\times10^{-22}$ to 1.9~$\times10^{-18}$ eV. We set limits on the coupling strength to baryon minus lepton number for each dark matter candidate reaching a peak sensitivity of $g_{B-L} \leq 9 \times 10^{-26}$.
\end{abstract}

\maketitle

\textit{Introduction} - While evidence for dark matter is extensive, direct detection has eluded all searches to date \cite{bertone2010particle, arcadi2018waning}. The lack of detection leaves open a wide parameter space of dark matter candidates including atypical scenarios that are not probed by current experiments.

Ultra-light dark matter (ULDM) has been suggested as an under-explored candidate \cite{graham2016dark}. These candidate particles can have masses ranging from $10^{-22}$ - 1 eV. ULDM searches have been conducted or proposed with a variety of techniques including atom interferometers \cite{PhysRevA.110.033313}, torsion balances \cite{shaw2022torsion, PhysRevD.111.063064}, and gravitational wave interferometers \cite{PhysRevD.110.042001}.

If dark matter is entirely made of such particles, then the observed dark matter density in the Milky Way would be enough for the ULDM to act similarly to a classical field \cite{hui2021wave}. This proposed ULDM field would have a characteristic frequency of:
\begin{equation}
f_{DM} \simeq \frac{m_{DM} c^2}{h}, 
\end{equation}
where $m_{DM}$ is the mass of the ULDM particle, $c$ is the speed of light, and $h$ is Planck's constant.

One proposed ULDM candidate is a vector particle that couples to the difference of baryon and lepton numbers ($B-L$) \cite{graham2016dark}. An ULDM field of this nature would interact with a pair of test bodies to produce a differential acceleration:
\begin{equation}
\Delta \vec{a} \simeq A \ g_{B-L}\ \Delta_{B-L}\ \hat{a},
\end{equation}
where $A= 4.9 \times 10^{11}\ \text{m/s}^2$ is a reference acceleration dependent on the local dark matter density \cite{graham2016dark}, $g_{B-L}$ is the ULDM coupling to $B-L$, $\Delta_{B-L}$ is the difference in baryon minus lepton number of the test bodies, and $\hat{a}$ is the direction of the ULDM vector field.

\begin{figure}[!h]
\includegraphics[width=0.5\textwidth]{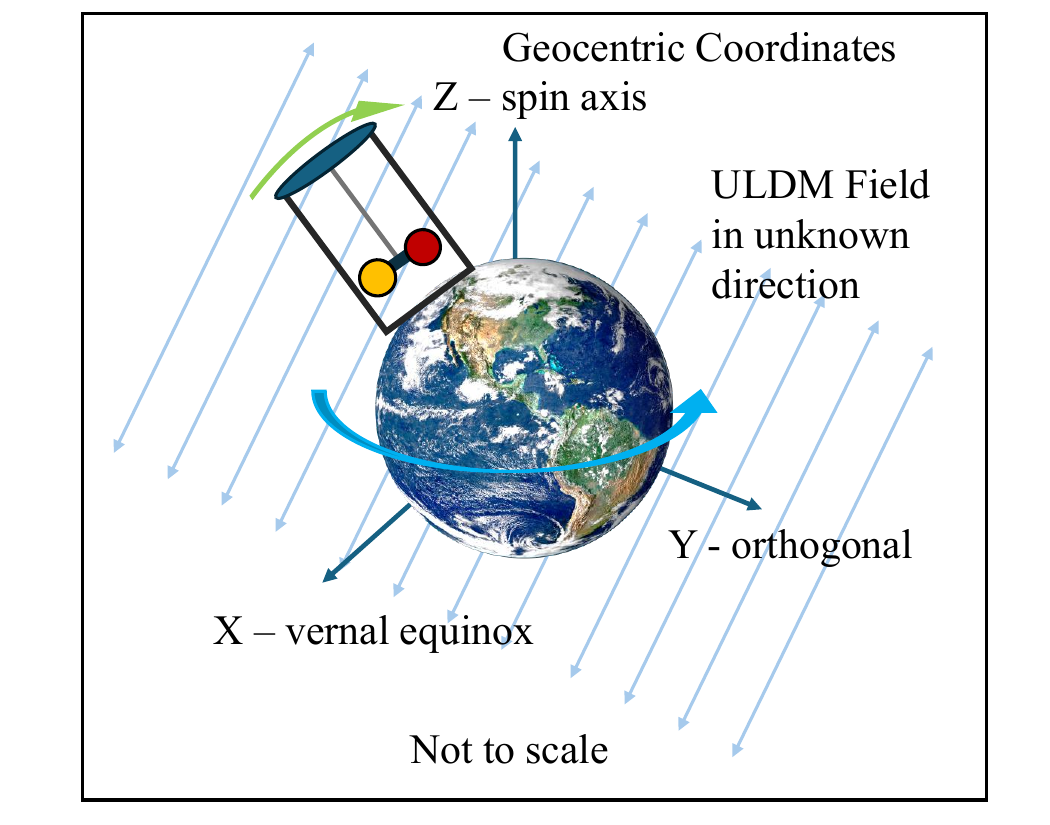}        
\caption{Diagram of the experiment showing the rotating torsion balance on the rotating Earth which is embedded in an ULDM field of unknown direction. The Z-component of the ULDM signal is modulated by the field's characteristic frequency and the rotation of the turntable while the X and Y components are additionally modulated by the Earth's rotation. Adapted from Reto St\"ockli, Nazmi El Saleous, and Marit Jentoft-Nilsen, NASA GSFC.}
\label{cartoon}
\end{figure}

We searched for this ULDM vector field using a rotating torsion balance with two different materials which formed a $B-L$ dipole. An ULDM field would exert a differential acceleration on the two materials and produce a measurable torque on the torsion balance.

\begin{figure}[!h]
\includegraphics[width=0.45\textwidth]{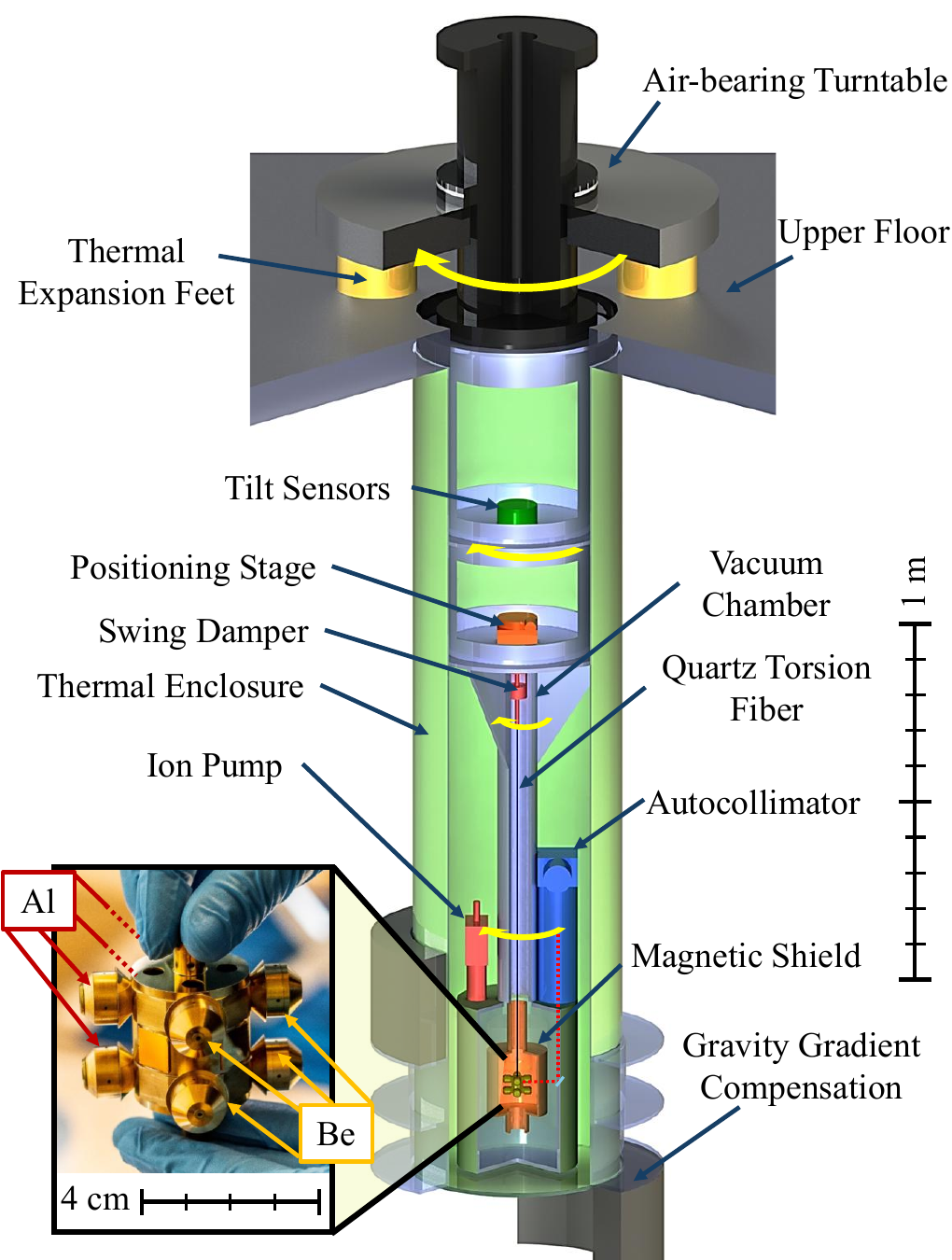}        
\caption{Schematic of the rotating torsion balance apparatus. The torsion pendulum, vacuum system, and angular readout are mounted on a rotating air-bearing turntable. The tilt of the apparatus is measured by a pair of co-rotating tilt sensors and controlled by thermal expansion feet that the turntable rests on. Thermal gradients and changes of the temperature, magnetic field, and gravity gradients are controlled, shielded, and compensated for, respectively. Reprinted from Ref. \cite{ross2025probing}.}
\label{fig:figures}
\end{figure}

\textit{Apparatus} - The torsion balance, shown in Fig. \ref{fig:figures}, was located at the Center for Experimental Nuclear Physics and Astrophysics on the University of Washington campus. The experiment was primarily operated to search for novel long-range interactions between normal matter and dark matter. However, the same data set used in Ref. \cite{ross2025probing} can be used to search for ULDM. 

The torsion balance consisted of a pendulum with eight test bodies, four made of beryllium and the other four of aluminum, in a dipole orientation. The pendulum was suspended from a 22-{\textmu}m thick, 1-m long fused silica fiber as described in Ref. \cite{shaw2023equivalence}. The relevant apparatus parameters are tabulated in Table \ref{appTable}. The balance was placed inside a vacuum chamber held at $<~0.2$ mPa maintained by an ion pump. The angle of the pendulum relative to the vacuum chamber was measured by an autocollimator. The performance of the apparatus was limited by thermal noise below 1 mHz equating to $\sim$0.5 fNm/$\sqrt{\text{Hz}}$ at 0.46 mHz \cite{thermal}.

\begin{table}[!h]

\begin{tabular}{ | m{0.2\textwidth} | m{0.05\textwidth}|| m{0.2\textwidth}|}
 \hline
 \multicolumn{3}{|c|}{Apparatus} \\
 \hline
 \vspace{0.02in}
 Parameter &  & Value \\
 \hline
 \vspace{0.05in}
 Spring constant & \centering{$\kappa$}  &\ $ 7.1 \times 10^{-10}$ N m/rad\\
 Quality factor & \centering{$Q$}  &\ $ 1.1\times 10^{5} $\\
 Mass of test body set & \centering{$m$}  &\ $  19.4 $ g\\
 Moment of inertia & \centering{$I$}  &\ $  3.78 \times 10^{-5}$ kg $\text{m}^2$\\
 Resonant frequency & \centering{$\omega_0$}  &\ $  2\pi \times 0.69$ mHz\\
 Lever-arm & \centering{$r$}  &\ $  1.9 $ cm\\
 Fiber length & \centering{$l$}  &\ $  0.94 $ m\\
Charge difference & \centering{$\Delta_{B-L}$}  &\ $  0.0359 $ \\
Rotation rate & \centering{$\omega_{TT}$}  &\ $  0.46 $ mHz\\
 \hline
\end{tabular}
 \caption{Relevant parameters of the torsion balance. The lever-arm $r$ is the radial distance from the pendulum's vertical axis to the center of mass of each set of test bodies.}\label{appTable}
\vspace{-0.2in}
\end{table}

The entire apparatus was supported by an air-bearing turntable which rotated at 0.46 mHz. The rotation rate was maintained with a FPGA controller that kept the variations in rotation rate to below 0.1 nrad/$\sqrt{\text{Hz}}$. Rotating the apparatus modulated the orientation of the composition dipole relative to the ULDM signal. This modulation allowed us to search for ULDM candidates at lower masses than an equivalent static torsion balance \cite{shaw2022torsion}. The tilt of the turntable was measured by a pair of co-rotating tiltmeters. The tilt signals were fed back to a set of thermal tilt compensation legs which held the tilt constant to a precision of 0.5~{\textmu}rad/$\sqrt{\text{Hz}}$.

The measured pendulum angle was converted to torque using the harmonic response of the torsion balance. The absolute calibration of the measured torque was done in three ways: with known parameters of the apparatus, with a speed change of the turntable, and with an injected gravitational torque. Further details can be found in Ref. \cite{ross2025probing}. 

\begin{figure}[!h]
\centering \includegraphics[width=0.45\textwidth]{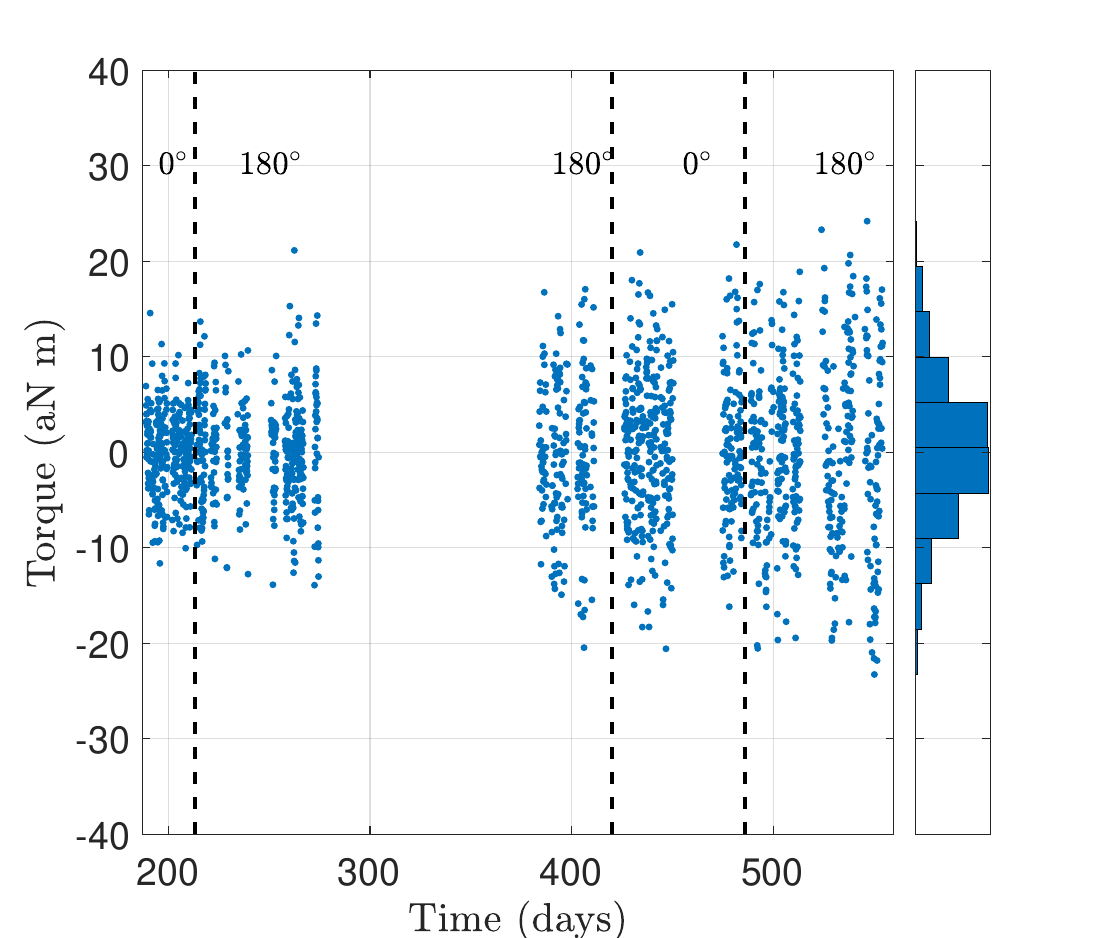}
\caption{Measured once per revolution torque amplitude over time. The zero time is set to January 1, 2024. Nearby construction activity and hardware failures caused many days of data to be dropped. However, at times when the construction activity was paused (weekends, holidays, etc.) the instrument was thermal noise limited.}
\label{time} 
\end{figure}

\textit{Analysis} - The experiment ran from July 7, 2024 to July 7, 2025. The pendulum was rotated $180^\circ$ on July 31, 2024,  February 15, 2025, and May 1, 2025 to minimize systematic effects. The operation of the apparatus was severely restricted due to noise from an active construction site $<$~100 m away as well as a series of hardware failures. The measured torque was split into segments each the length of two turntable revolution periods. Each segment was independently fit to a series of sinusoidal functions with frequencies of harmonics of the turntable frequency ($\omega_{TT}$, $2\omega_{TT}$, $3\omega_{TT}$, etc.), and harmonics of the torsional resonance ($\omega_0$, $2\omega_0$, $3\omega_0$, etc.). 

The instrument experienced frequent transients due to earthquakes, construction activity, and other environmental disturbances. We discarded data points that did not fall under two criteria: a misfit-squared exceeding the 99th
percentile of the corresponding $\chi^2$-distribution and a torque amplitude outside of the 95\% interval. 
\pagebreak

\begin{widetext}    

\begin{figure}[!t]
\centering \includegraphics[width=1\textwidth]{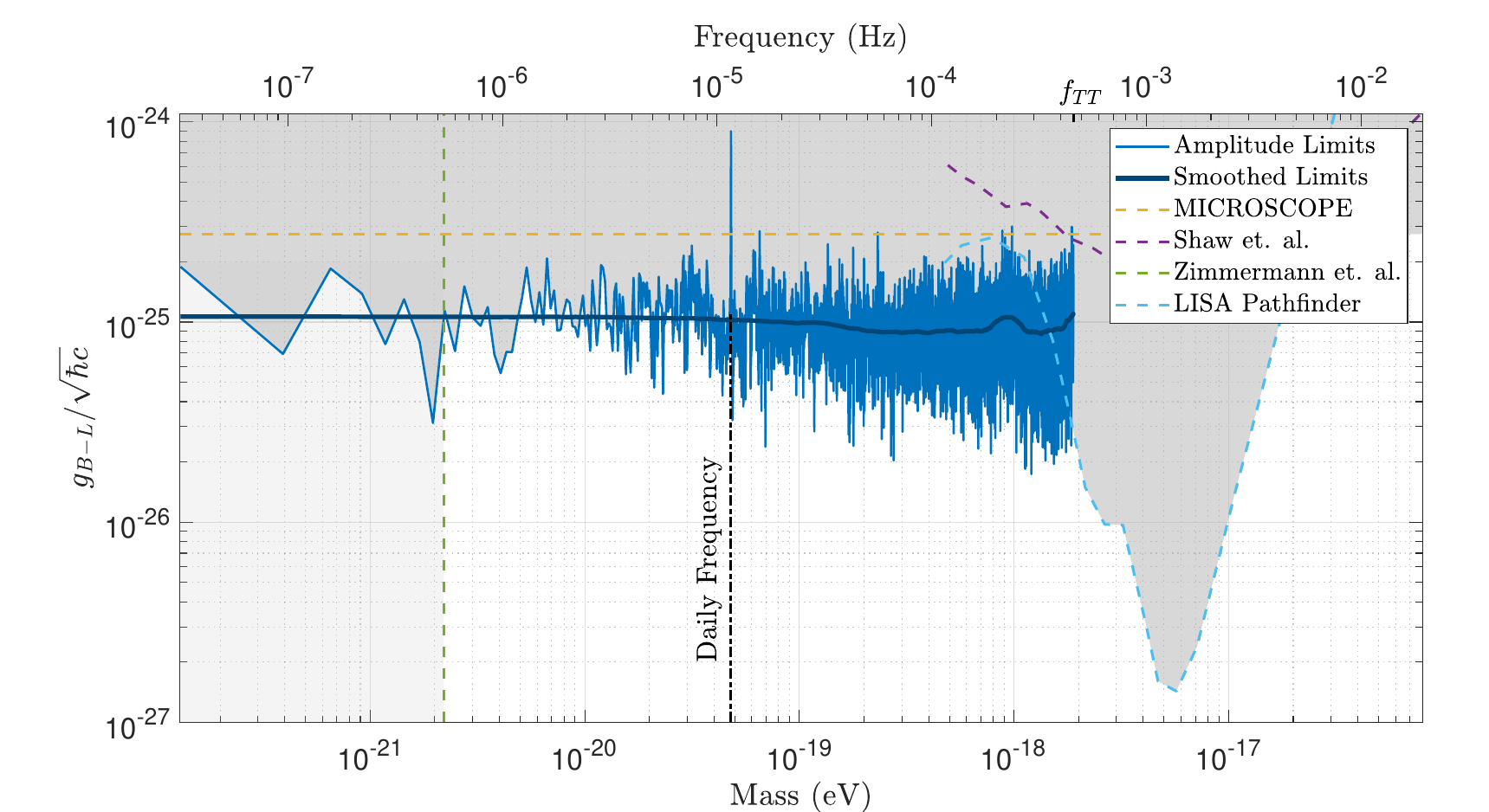}
\caption{Our 95\%-confidence upper limits on the coupling constant of ULDM to B-L, $g_{B-L}$, along with previous limits set by MICROSCOPE \cite{frerick2024riding}, \citet{shaw2022torsion}, \citet{PhysRevLett.134.151001}, LISA Pathfinder \cite{frerick2024riding}, and the combined best direct detection limits.}
\label{limits} 
\end{figure}

\end{widetext} 

Fig. \ref{time} shows the surviving (68\% of full data set) once per turntable revolution torque amplitudes, which are consistent with a Gaussian distribution. We searched for ULDM interacting with the $B-L$ dipole of the pendulum which was modulated at the turntable frequency. The once per turntable revolution amplitude was split into 30 day long sections. Each section was then fit to a set of basis functions representing an alternating ULDM field with a characteristic frequency pointing along either the $X$, $Y$, or $Z$ directions.
These directions are defined in a geocentric coordinate system with the $Z$ along the Earth's spin axis, $X$ pointing in the direction of the vernal equinox, and $Y$ orthogonal to $X$ and $Z$ as shown in Fig. \ref{cartoon}.

These unit vectors were then projected into the laboratory frame using the {\tt Astropy} libraries \cite{astropy:2022}.

The observed torque was fit to:
\begin{equation}
    \tau = A\ m\ r\Delta_{B-L}\ \bigg( g_{B-L}^X \Vec{X} \cdot \Vec{p} + g_{B-L}^Y \Vec{Y} \cdot \Vec{p} + g_{B-L}^Z \Vec{Z} \cdot \Vec{p}\bigg)
\end{equation}
where $g_{B-L}^{X,Y,Z}$ are respectively the couplings to an ULDM field along $X$, $Y$, and $Z$, and $\Vec{p}$ is the horizontal vector normal to the dipole of the pendulum. The amplitudes in each direction of a given mass were then averaged in quadrature to yield a mass-dependent limit on the ULDM coupling.

\textit{Results} - The 95\%-confidence upper limits on the ULDM coupling to B-L for the mass range of 1.3 $\times10^{-22}$ to 1.9 $\times10^{-18}$ eV with mass spacing of 2.6 $\times10^{-22}$ eV are shown in Fig. \ref{limits}. The mass range was restricted by the length of the data set on the low mass end and the turntable frequency on the high mass end.

The limits do not exhibit strong frequency dependence. The uncertainty is primarily driven by the thermal noise of the torsion balance at the turntable frequency and the length of the data campaign. The peak at 4.8 $\times10^{-20}$ eV corresponds to the daily frequency. The experiment can not distinguish a signal from a dark matter candidate at this mass from the modulation due to the Earth's rotation.

\textit{Conclusion} - We searched for ULDM candidates coupled to a B-L dipole. We find no evidence of a signal from ULDM and set 95\%-confidence upper limits on the coupling strength of ULDM to B-L at peak sensitivity of $g_{B-L}< 9 \times 10^{-26}$. These limits improve upon previous limits by a factor of three and extend the reach of torsion balance searches of dark matter. 

This experiment restricts the possible candidates for dark matter at the ultra low end of the possible mass spectrum. With further upgrades (e.g. ULDM specific pendulum) this torsion balance apparatus can extend further into unexplored dark matter parameter space.

\textit{Data Availability} - Code and data to generate the results shown here can be found at:\\
\url{https://github.com/EotWash/ULDM_NewWash}\\ Raw data available upon request.

\textit{Acknowledgments} - This work was supported by funding from the NSF under Awards PHY-1607385, PHY-1607391, PHY-1912380, and PHY-1912514.

\bibliographystyle{unsrtnat}
\bibliography{main.bib}

\begin{thebibliography}{14}
\providecommand{\natexlab}[1]{#1}
\providecommand{\url}[1]{\texttt{#1}}
\expandafter\ifx\csname urlstyle\endcsname\relax
  \providecommand{\doi}[1]{doi: #1}\else
  \providecommand{\doi}{doi: \begingroup \urlstyle{rm}\Url}\fi

\bibitem[Bertone(2010)]{bertone2010particle}
Gianfranco Bertone.
\newblock \emph{Particle dark matter: observations, models and searches}.
\newblock Cambridge University Press, 2010.

\bibitem[Arcadi et~al.(2018)Arcadi, Dutra, Ghosh, Lindner, Mambrini, Pierre, Profumo, and Queiroz]{arcadi2018waning}
Giorgio Arcadi, Ma{\'\i}ra Dutra, Pradipta Ghosh, Manfred Lindner, Yann Mambrini, Mathias Pierre, Stefano Profumo, and Farinaldo~S Queiroz.
\newblock The waning of the wimp? a review of models, searches, and constraints.
\newblock \emph{The European Physical Journal C}, 78:\penalty0 1--57, 2018.

\bibitem[Graham et~al.(2016)Graham, Kaplan, Mardon, Rajendran, and Terrano]{graham2016dark}
Peter~W Graham, David~E Kaplan, Jeremy Mardon, Surjeet Rajendran, and William~A Terrano.
\newblock Dark matter direct detection with accelerometers.
\newblock \emph{Physical Review D}, 93\penalty0 (7):\penalty0 075029, 2016.

\bibitem[Zhou et~al.(2024)Zhou, Ranson, Panagiotou, and Overstreet]{PhysRevA.110.033313}
Yifan Zhou, Rowan Ranson, Michalis Panagiotou, and Chris Overstreet.
\newblock Ytterbium atom interferometry for dark matter searches.
\newblock \emph{Phys. Rev. A}, 110:\penalty0 033313, Sep 2024.
\newblock \doi{10.1103/PhysRevA.110.033313}.
\newblock URL \url{https://link.aps.org/doi/10.1103/PhysRevA.110.033313}.

\bibitem[Shaw et~al.(2022)Shaw, Ross, Hagedorn, Adelberger, and Gundlach]{shaw2022torsion}
EA~Shaw, MP~Ross, CA~Hagedorn, EG~Adelberger, and JH~Gundlach.
\newblock Torsion-balance search for ultralow-mass bosonic dark matter.
\newblock \emph{Physical Review D}, 105\penalty0 (4):\penalty0 042007, 2022.

\bibitem[Sun et~al.(2025)Sun, Slagmolen, and Qin]{PhysRevD.111.063064}
Ling Sun, Bram J.~J. Slagmolen, and Jiayi Qin.
\newblock Differential torsion sensor for direct detection of ultralight vector dark matter.
\newblock \emph{Phys. Rev. D}, 111:\penalty0 063064, Mar 2025.
\newblock \doi{10.1103/PhysRevD.111.063064}.
\newblock URL \url{https://link.aps.org/doi/10.1103/PhysRevD.111.063064}.

\bibitem[Abac et~al.(2024)]{PhysRevD.110.042001}
A.~G. Abac et~al.
\newblock Ultralight vector dark matter search using data from the kagra o3gk run.
\newblock \emph{Phys. Rev. D}, 110:\penalty0 042001, Aug 2024.
\newblock \doi{10.1103/PhysRevD.110.042001}.
\newblock URL \url{https://link.aps.org/doi/10.1103/PhysRevD.110.042001}.

\bibitem[Hui(2021)]{hui2021wave}
Lam Hui.
\newblock Wave dark matter.
\newblock \emph{Annual Review of Astronomy and Astrophysics}, 59\penalty0 (1):\penalty0 247--289, 2021.

\bibitem[Ross et~al.(2025)Ross, Shaw, Gettings, Apple, Paulson, and Gundlach]{ross2025probing}
MP~Ross, EA~Shaw, C~Gettings, SK~Apple, IA~Paulson, and JH~Gundlach.
\newblock Probing for non-gravitational long-range dark matter interactions.
\newblock \emph{arXiv preprint arXiv:2509.10701}, 2025.

\bibitem[Shaw(2023)]{shaw2023equivalence}
Erik~A Shaw.
\newblock \emph{Equivalence Principle Tests and Direct Searches for Ultra-Light Dark Matter with Fused-Silica Torsion Fibers}.
\newblock University of Washington, 2023.

\bibitem[Saulson(1990)]{thermal}
Peter~R. Saulson.
\newblock Thermal noise in mechanical experiments.
\newblock \emph{Phys. Rev. D}, 42:\penalty0 2437--2445, Oct 1990.
\newblock \doi{10.1103/PhysRevD.42.2437}.
\newblock URL \url{https://link.aps.org/doi/10.1103/PhysRevD.42.2437}.

\bibitem[Frerick et~al.(2024)Frerick, Jaeckel, Kahlhoefer, and Schmidt-Hoberg]{frerick2024riding}
Jonas Frerick, Joerg Jaeckel, Felix Kahlhoefer, and Kai Schmidt-Hoberg.
\newblock Riding the dark matter wave: Novel limits on general dark photons from lisa pathfinder.
\newblock \emph{Physics Letters B}, 848:\penalty0 138328, 2024.

\bibitem[Zimmermann et~al.(2025)Zimmermann, Alvey, Marsh, Fairbairn, and Read]{PhysRevLett.134.151001}
Tim Zimmermann, James Alvey, David J.~E. Marsh, Malcolm Fairbairn, and Justin~I. Read.
\newblock Dwarf galaxies imply dark matter is heavier than $2.2\ifmmode\times\else\texttimes\fi{}{10}^{\ensuremath{-}21}\text{ }\text{ }\mathrm{eV}$.
\newblock \emph{Phys. Rev. Lett.}, 134:\penalty0 151001, Apr 2025.
\newblock \doi{10.1103/PhysRevLett.134.151001}.
\newblock URL \url{https://link.aps.org/doi/10.1103/PhysRevLett.134.151001}.

\bibitem[{Astropy Collaboration} and {Astropy Project Contributors}(2022)]{astropy:2022}
{Astropy Collaboration} and {Astropy Project Contributors}.
\newblock {The Astropy Project: Sustaining and Growing a Community-oriented Open-source Project and the Latest Major Release (v5.0) of the Core Package}.
\newblock \emph{\apj}, 935\penalty0 (2):\penalty0 167, August 2022.
\newblock \doi{10.3847/1538-4357/ac7c74}.

\end{thebibliography}

\end{document}